\documentclass[aps,prb,twocolumn,superscriptaddress,showpacs,floatfix]{revtex4}
\usepackage{graphicx}
\usepackage{amssymb,amsmath}
\newcommand{\btau}{\mbox{\boldmath$\tau$}}

\begin{document}


\title{Stripeless incommensurate magnetism in strongly correlated oxide La$_{1.5}$Sr$_{0.5}$CoO$_{4}$}


\author{A.~T.~Savici}
\author{I.~A.~Zaliznyak}
\author{G.~D.~Gu}
\affiliation{DCMPMS,
Brookhaven National Laboratory, Upton, New York 11973-5000}
\author{R.~Erwin}
\affiliation{NCNR, National Institute of Standards and Technology,
Gaithersburg, Maryland 20899, USA}


\date{\today}

\begin{abstract}
We studied the nano-scale structure of the short-range
incommensurate magnetic order in La$_{1.5}$Sr$_{0.5}$CoO$_{4}$ by
elastic neutron scattering. We find that magnetic diffuse scattering
is isotropic in the $a-b$ plane, in contrast with the naive
expectation based on the popular stripe model. Indeed, charge
segregation into lines favoring certain lattice direction(s) would
facilitate linear stacking faults in an otherwise robust
antiferromagnetism of un-doped material, leading to anisotropic
disorder, with a characteristic symmetry pattern present in the
neutron scattering data.

\end{abstract}

\pacs{
71.28+d,
75.40.Cx,
75.40.Gb,
75.50.Ee}

\maketitle


\section{Introduction}

Ever since the advent of high-temperature superconductivity (HTSC)
in cuprates, the physics of doped strongly correlated transition
metal oxides remains at the forefront of condensed matter research.
\cite{LeeNagaosaWen2006,OrensteinMillis2000,Tokura2003} In
particular, there is a renewed interest in metal-insulator
transitions associated with charge/orbital ordering in doped
manganese and nickel oxides and in "colossal" magnetoresistance
phenomena. \cite{Tokura2003,TokuraNagaosa2000,Imada1998} While
macroscopic magnetic and transport properties of strongly correlated
oxides respond to doping in many different and often fascinating
ways, the appearance of structural and magnetic superlattices whose
periods depend on the doping level is a common microscopic response
shared by many oxides.
\cite{Tranquada1994,Tranquada1995,Wells1997,Chen1993,Yoshizawa2000,Kajimoto2001,Ishizaka2003,Larochelle2001,Sakiyama2006,Zaliznyak}

Simultaneous incommensurate magnetic and charge ordering was
probably first observed in a doped nickelate,
La$_{2-x}$Sr$_x$NiO$_{4+y}$. \cite{Tranquada1994} It gained
prominence when a similar phenomenon was discovered in a $x \approx
1/8$ doped cuprate with an anomalously suppressed superconductivity.
\cite{Tranquada1995} It was proposed that a simple model of
real-space static ordering of holes and spins, where doped charges
segregate into lines separating magnetically ordered stripe domains,
can explain all features observed by elastic neutron scattering. In
conjunction with earlier theoretical predictions of such
superstructures in the 2D Hubbard model, which is believed to
describe the physics of HTSC cuprates,
\cite{Zaanen1989,Emery1993,Machida1989,Machida1990} striped phases
gained broad popularity and became essentially a default model for
describing incommensurate magnetic and charge superstructures in
doped layered perovskite oxides La$_{2-x}$Sr$_x$MO$_4$ (M = Cu, Ni,
Co, Mn).

There is a growing recognition, however, that physics of charge
ordering in cuprates may differ significantly from that in
well-insulating materials such as cobaltates and nickelates, where
it can also be viewed as ordering of polarons driven by lattice
elastic interactions. \cite{Zaanen1994,Khomskii2001,Collart2006} In
fact, it was argued theoretically that formation of superstructures
whose period depends on the doping level, including stripes, is a
natural response of the crystal lattice to local strain associated
with doped charges and can be already explained by considering the
system's elastic energy. \cite{Khomskii2001} Experiments indicate
this type of superlattices in layered manganates and cobaltates.
\cite{Sakiyama2006,Zaliznyak,Larochelle2001}

\begin{figure}[!b]
\begin{center}
\includegraphics[width=3.2in,angle=0]{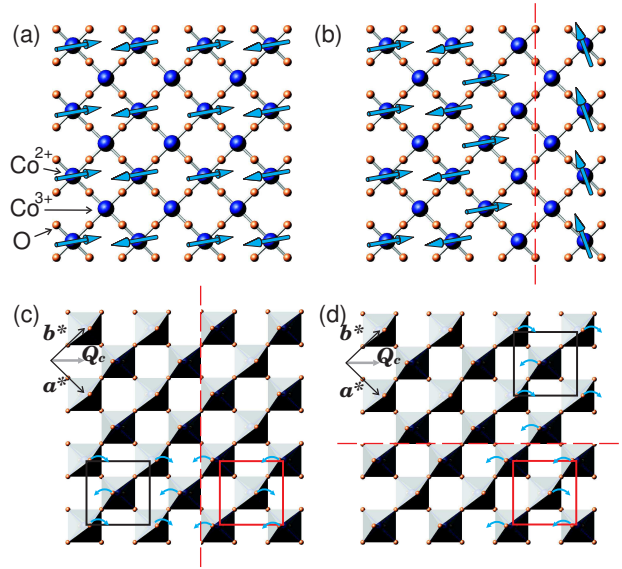}%
\caption{\label{superlattice} (a) Checkerboard charge and spin order at
half-doping. (b) Stacking fault giving rise to short-range correlation
and magnetic incommensurability in La$_{1.5}$Sr$_{0.5}$CoO$_4$ in
stripe picture. (c), (d) LTO superlattice of weakly doped cuprates.
${\bf a^*}$, ${\bf b^*}$ are reciprocal lattice vectors of the HTT
phase, arrows show tilts of O octahedra. Stacking faults separating
structural domains with opposite tilts (broken lines) running along
"stripes" (perpendicular to $Q_c$), (c), and perpendicular to "stripes"
(parallel to $Q_c$), (d), have the same energy, implying isotropic
disorder. }
\end{center}
\end{figure}

In a superlattice, e.g. such as associated with polaron ordering,
atomic positions and/or alignment of magnetic moments do not vary in
the direction perpendicular to the propagation vector, presenting
superlattice modulation as a periodic arrangement of lines of
parallel spins and/or identical atomic displacements (Fig.
\ref{superlattice}). Hence, stripe superstructures resulting from
one-dimensional electronic phase segregation and elastic/magnetic
superlattices have similar appearance in real space. Accounting for
domains, they also give rise to similarly positioned elastic peaks
observable in scattering experiments. Hence, the question arises: is
it possible to distinguish between the two cases? Here we show that
for short-range superstructures this question can be answered by
studying the nano-scale structure of disorder. By measuring the
pattern of elastic neutron scattering, we find that short-range
incommensurate magnetism in half-doped cobaltite
La$_{1.5}$Sr$_{0.5}$CoO$_4$ does not originate from an intrinsically
one-dimensional stripe charge order.

At half-doping, the system is naturally amenable to a checkerboard
charge order (CO) where every other site in the $a-b$ plane of the
high-temperature tetragonal (HTT) structure accommodates a hole,
Fig. \ref{superlattice} (a). It is accompanied by a correlated
harmonic modulation of atomic positions with propagation vector $Q_c
= (1/2,1/2)$ in the HTT reciprocal lattice, resulting in a
superlattice with twice larger unit cell compared to the HTT phase
where $a = b \approx 3.83$ \AA. In stripe picture this type of CO
can be viewed as an alternate stacking of diagonal charge stripes.
The CO structural disorder results from faults in stripe stacking,
Fig. \ref{superlattice} (b), and is one-dimensional (1D) in nature.
The ordering of small polarons driven by lattice strain, on the
other hand, is in essence similar to the cooperative tilt pattern of
oxygen octahedra in the low-temperature orthorhombic (LTO) lattice,
which relieves chemical pressure in weakly doped cuprates, Fig.
1(c,d). Stacking faults have no intrinsic 1D rigidity and result in
randomly shaped domains and isotropic disorder.

Although $x=0.5$ regime is inaccessible in cuprates, checkerboard CO
was found for M = Ni, Co, Mn.
\cite{Kajimoto2001,Larochelle2001,Sakiyama2006,Zaliznyak} While
there is no yet consensus on the cobaltate, ordering in Mn material
is commonly viewed as a cooperative Jahn-Teller distortion, or CDW,
driven mainly by lattice elastic energy, \cite{Larochelle2001} while
that in the nickelate is usually discussed in terms of stripes,
\cite{Kajimoto2001} following the original proposition of Ref.
\onlinecite{Tranquada1994}. Although for different reasons, hole
sites are effectively nonmagnetic both in cobaltite and nickelate;
antiferromagnetic spin order (SO) on the remaining sites gives rise
to a superlattice with four times the period of the original HTT
lattice, Fig. 1(a). Experiments show that this spin order is usually
short-ranged, most probably reflecting the short-range nature of
charge/stripe superlattice. Then, it would be natural to expect that
structure of these short-range nano-scale spin correlations reflects
the structure of faults in the charge order, e. g. a disorder in the
form of linear magnetic disclinations associated with stripe
stacking faults, Fig. 1(b).

\section{Experimental procedure}

We studied a large single crystal of La$_{1.5}$Sr$_{0.5}$CoO$_{4}$
($m\approx 11$ g) grown by the floating zone method. It has a nearly
HTT structure with lattice parameters $a = b \approx 3.83$ \AA\ and
$c \approx 12.5$ \AA\ at T = 10 K and was previously described in
Ref. \onlinecite{Zaliznyak}. Sample mosaic spread is $\eta \approx
20'$. Measurements were done in (h,k,0) and (h,h,l) reciprocal
lattice zones using cold (SPINS) and thermal (BT9) neutron triple
axis spectrometers, respectively, at NIST Center for Neutron
Research. Monochromatic neutrons were obtained using (002)
reflection from vertically focussing pyrolytic graphite (PG)
crystals and analyzed using flat PG(002) analyzer crystals. On SPINS
beam collimations were $\approx 37'-80'-80'-240'$, from guide to
detector, and neutron final energy was $E_f = 5$ meV. Beryllium
filters both before and after the sample were used to remove the
contamination from higher order reflections in PG. On BT9 we used
$E_f = 14.7$ meV, collimations $\approx 40'-40'-40'-100'$, and PG
filters before and after the sample.

\begin{figure}[!b]
\begin{center}
\includegraphics[width=3.2in,angle=0]{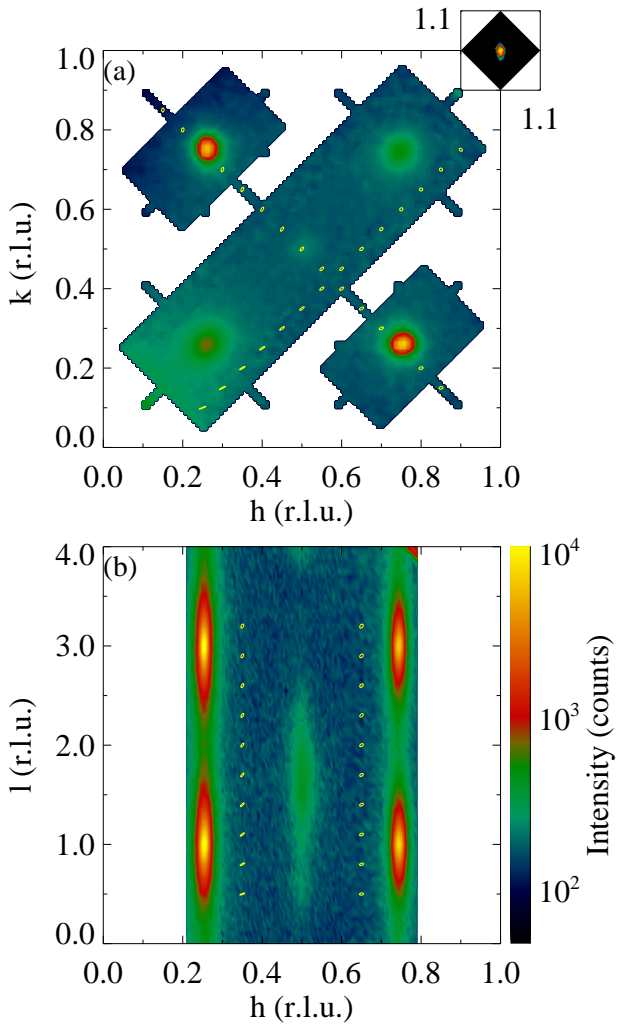}%
\caption{\label{data} Contour map of the measured neutron elastic
scattering intensity in (h,k,0) (a) and (h,h,l) (b) zones at T = 3.5
K and 10 K, respectively. Ellipses show the calculated full width at
half maximum (FWHM) instrument resolution.\cite{Cooper-Nathans}
Magnetic peaks are at $h,k \approx 0.25,0.75$ r.l.u. Charge order
scattering is seen at h = k = 0.5 r.l.u. Intensity in the map around
(1,1,0) Bragg peak shown in the top right corner was scaled down by
a factor of 100.}
\end{center}
\end{figure}

Color contour maps of the measured elastic scattering intensity are
shown in Fig. \ref{data} (a,b). Both in (h,k,0) (a) and (h,h,l) (b)
zones the observed peaks are much broader than the calculated
instrument resolution, which is illustrated by the FWHM ellipses.
Peaks of magnetic origin are at $h, k \approx 0.25$ and 0.75, while
those due to atomic displacement accompanying charge ordering are at
h = k = 0.5. Checkerboard CO in La$_{1.5}$Sr$_{0.5}$CoO$_{4}$ sets
in at about 825 K, while magnetic spin ordering appears only below
abound 30 K. \cite{Zaliznyak}

 \begin{figure}[!t]
 \includegraphics[width=3.2in,angle=0]{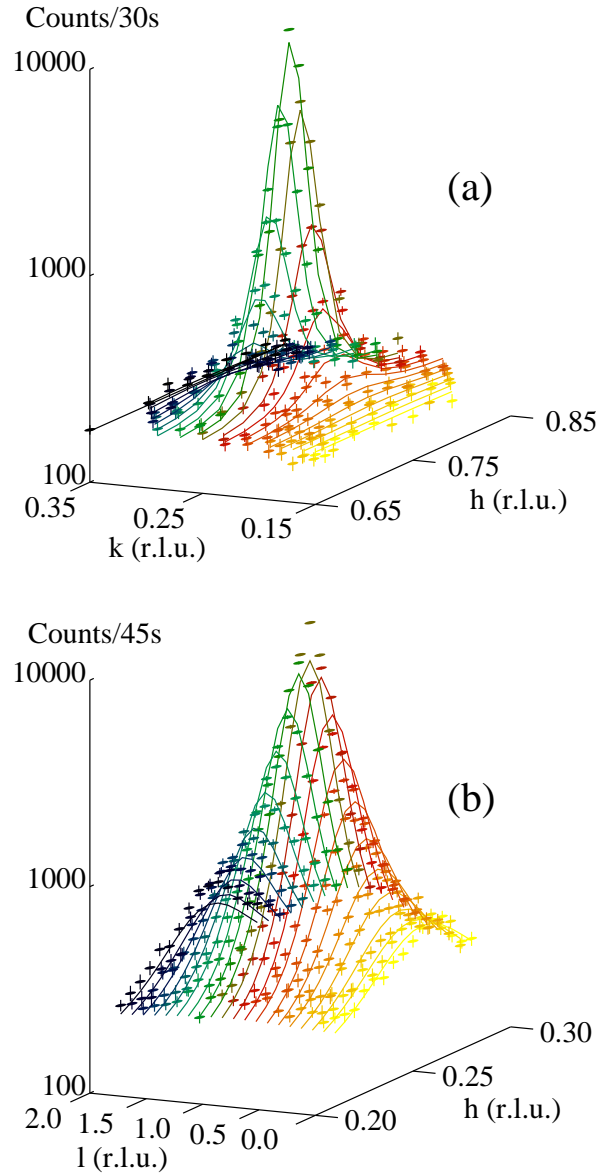}
 \caption{\label{scans} Elastic neutron scattering from
La$_{1.5}$Sr$_{0.5}$CoO$_4$. (a) (h,k,0) reciprocal lattice zone,
(b) (h,h,l) zone. The lines show the global fit of the data to Eq.
(\ref{ndim}) describing coupled anisotropic 3D correlations.}
 \end{figure}

Selected scans around the magnetic peak position are presented in
Figure \ref{scans}. The lines show the result of the global fit of
all data to the resolution corrected anisotropic cross section given
by Eq. (\ref{ndim}) for $D = 3$, which is discussed in detail in the
next section.

\begin{figure}[!h]
\begin{center}
\includegraphics[width=3.2in,angle=0]{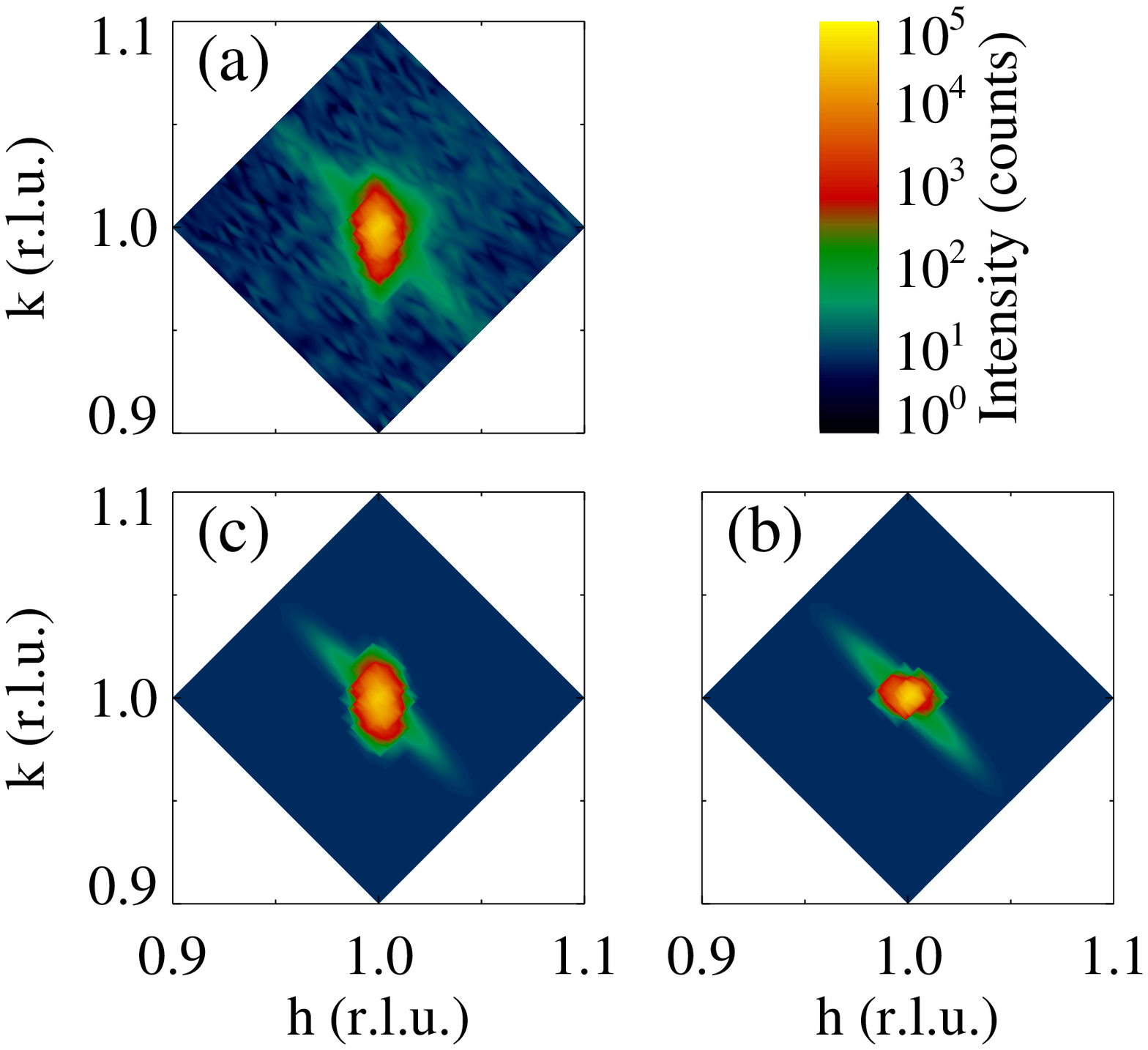}%
\caption{\label{Bragg110} Resolution and sample size effects for
(1,1,0) Bragg reflection. (a) Contour map of the measured neutron
scattering intensity in the (h,k,0) zone at T = 3.5 K. (b)
Calculated intensity for the point-like sample, corresponding to the
resolution-broadened delta function.\cite{Cooper-Nathans} (c) Same
as in (b), but with sample size effects included. Small extra
intensity extended along a diagonal corresponds to a small ($\sim
0.3\%$) oriented powder component with angular distribution of
$\approx 0.8^\circ$, much broader than sample mosaic spread $\eta
\approx 0.3^\circ$. }
\end{center}
\end{figure}

Quantifying the nano-scale structure of short-range magnetic
correlations experimentally so as to distinguish between various
symmetries of magnetic domains requires accurate knowledge and deep
understanding of the resolution effects present in neutron
scattering measurements. In order to quantify the resolution
function and accurately account for the resolution effects we
measured elastic scattering intensity around (1,1,0) Bragg peak, in
the (h,k,0) orientation. The data scaled down by a factor of 100 in
order to roughly fit in the intensity range of magnetic scattering
are shown in the upper right corner of the contour plot of Figure
\ref{data}(a). It is immediately clear that magnetic peaks are much
broader than nuclear lattice peak, whose width is governed entirely
by the resolution and sample mosaics. A more detailed image of
intensity around (1,1,0) nuclear Bragg peak is shown in Fig.
\ref{Bragg110} (a). Intensity scale enhances regions with smaller
number of counts away from the peak.

Resolution of the triple axis neutron spectrometer is usually
described using Cooper-Nathans formalism.\cite{Cooper-Nathans} We
show the result of such calculation for the (110) nuclear Bragg
reflection in Fig. \ref{Bragg110} (b). There is an obvious
discrepancy between the calculation and the observed intensity shown
in Fig. \ref{Bragg110} (a), which has a pronounced tail along
$k$-direction, resulting in an elliptical Bragg spot on the contour
map. This shape can be explained by taking into account the sample
size effects (our sample is a $\approx 5$ cm long cylinder, which
for scattering measurement in the (h,k,0) zone is mounted roughly
parallel to the scattering plane). In a very general way, this can
be done by using the method devised by Popovici, \cite{Popovici}
where sample is described in terms of a Gaussian density
distribution. Instead, here we have explicitly included the sample
size in the Cooper-Nathans calculation by averaging over the
scattering angle between the incident and scattered neutron beams,
which varies across the length of the sample. The result of such
calculation gives very good fit of the measured (110) nuclear Bragg
intensity, which is plotted in Fig. \ref{Bragg110} (c). Sample
mosaic was found to be $0.3^\circ$.

Mosaic structure of the (110) nuclear Bragg reflection in Fig.
\ref{Bragg110} (a) consists of a single peak, with no apparent
indication of an orthorhombic lattice distortion. \cite{Matsuura}
Therefore, if present, any such distortion is not detectable within
the accuracy of our measurement of nuclear Bragg reflections.
However, we find a small, $\sim 0.6\%$ distortion from the analysis
of magnetic scattering, which is described below.

\section{Analysis and discussion}

An appealing feature of stripe picture is that it provides a simple
real-space model explaining both temperature-dependent CO
incommensurability in nickelates and short-range incommensurate
magnetism in both Ni and Co materials. \cite{Kajimoto2001} In this
picture they arise from discommensurations, or faults in the
stacking pattern of 1D charge/spin stripes, favored by strong
nearest-neighbor exchange coupling on the HTT square lattice, Fig.
\ref{superlattice} (b). At half-doping such faults effectively
reduce the average period of magnetic structure within the
correlated domains in $a-b$ plane, consistent with slightly longer
than (1/4,1/4) SO wave vector $Q_{\rm so} \approx (0.256,0.256)$ in
La$_{1.5}$Sr$_{0.5}$CoO$_4$. \cite{Zaliznyak} It is also clear from
the figure that discommensurations introduce linear disclinations
parallel to stripes (coupling across two consecutive hole sites is
weak and frustrated) and therefore truncate spin correlation range.
This type of disorder has a specific imprint in the structure of
diffuse elastic peaks measured in scattering experiment.
\cite{Cowley,Wilson1942,Warren,Krivoglaz,Chen2005}

Anisotropic short-range-ordered superlattices are well known in the
physics of imperfect crystals and binary alloys, such as Cu$_{3}$Au.
\cite{Cowley,Wilson1942,Warren,Krivoglaz} Phase mismatches at the
boundaries of antiphase domains and/or stacking faults introduce
one-dimensional disorder in the direction perpendicular to the
defect planes. A combination of several systems of such phase slips
allowed in the crystal structure leads to a peculiar X-ray (and
neutron) scattering pattern, with tails along certain lattice
directions. \cite{Cowley,Warren}

Similar considerations can be extended to scattering by short-range
magnetic structures where disorder results from un-correlated
stacking faults (disclinations), such as shown in Fig.
\ref{superlattice} (b). \cite{ZaliznyakLee} The elastic magnetic
neutron scattering cross section is given by
\begin{equation}\label{magcs}
 \frac{d\sigma({\bf q})}{d\Omega} =
 N \left(\frac{r_{m}}{2\mu_{B}}\right)^{2} \sum_{j}^{N} {\rm e}^{-i {\bf q}\cdot{\bf R}_j}
 \langle {\bf M}_{0}^{\bot}(-{\bf q}){\bf M}_{j}^{\bot}({\bf
 q})\rangle,
\end{equation}

\noindent where $r_{m} \approx -5.39\cdot 10^{-13}$cm, $\mu_B$ is
the Bohr's magneton, ${\bf M}_{j}^{\bot}({\bf q})$ is the
perpendicular to the wave vector $\bf q$ component of the
Fourier-transform of the magnetization of atoms belonging to the
lattice unit cell at a position ${\bf R}_j$, and the sum extends
over all $N$ unit cells of the crystal. In the presence of
long-range magnetic order with wave vector {\bf Q},
\begin{equation}\label{longrange}
\langle {\bf M}_{j}({\bf q}) \rangle = {\bf m}({\bf q}) {\rm e}^{i {\bf
Q}\cdot{\bf R}_j} + {\bf m}^*({\bf q}) {\rm e}^{-i {\bf Q}\cdot{\bf
R}_j},
\end{equation}

\noindent where the order parameter ${\bf m}({\bf q})$ includes
Wannier function describing magnetic form factor of the unit cell.
Fourier-transform in Eq. (\ref{magcs}) is a sum of delta-functions
offset by $\bf Q$ from reciprocal lattice points.

Un-correlated magnetic disclinations in the crystal can be accounted
for by introducing additional random phase multipliers
$e^{-i\phi_{j}}$ in the magnetization density (\ref{longrange}). In
view of its randomness, averaging over this phase factor can be
decoupled in the correlation function in Eq. (\ref{magcs}). Assuming
self-averaging and Gaussian randomness, its statistical average is
$\langle e^{-i\phi_{j}}\rangle=e^{-\langle \phi_{j}^{2}\rangle /2}$
(Bloch identity) and the scattering cross-section is
\begin{equation}\label{sumlattice}
 \frac{d\sigma({\bf q})}{d\Omega} =
 N r_{m}^2 \left|\frac{{\bf m}^{\bot}({\bf q})}{2\mu_{B}}\right|^{2}
 \sum_{j}^{N} {\rm e}^{-i {\bf q}\cdot{\bf R}_{j}-\frac{1}{2} \langle \phi_{j}^{2}\rangle}.
\end{equation}

In the case of planar (linear in 2D) disclinations perpendicular to
principal lattice directions such as expected from stripes, the
accumulated mean-square phase mismatch can be described by
independent random walks along these directions. Then, $\langle
\phi_{j}^{2}\rangle/2 = \sum_{\alpha} |n_{j,\alpha}|/\xi_{\alpha}$,
where $n_{j,\alpha}$ label lattice sites, ${\bf
R}_{j}=\sum_{\alpha}n_{j,\alpha}{\bf a}_{\alpha}$, and
$\xi_{\alpha}$ are correlation lengths in appropriate units ($\alpha
= x,y,z$). Substituting this into Eq. (\ref{magcs}), one obtains
cross-section in the form of a product of 1D lattice-Lorentzians
(LL),
\begin{equation}\label{latticelorentzian}
 \tilde{L}_{\xi_{\alpha}}(q_{\alpha})\equiv \frac{\sinh \xi_{\alpha}^{-1}}{\cosh \xi_{\alpha}^{-1}-
 \cos(q_\alpha \pm Q_{\alpha})},
\end{equation}

\noindent along principal crystallographic directions (Eq.
(\ref{latticelorentzian}) is a sum of Lorentzians placed
periodically in reciprocal lattice). Factorized cross-section is a
consequence of the 1D nature of disorder generated by system of
linear/planar phase slips. It retains the orientational symmetry of
these defects in the crystal lattice.

 \begin{table}[!ht]
 \caption{\label{coupling}Scattering functions for different structure of the nano-scale
 disorder on a 3D lattice (assuming large $\xi_\alpha$).}
 \begin{ruledtabular}
 \begin{tabular}{lc}
 type of disorder & scattering cross section\\
 \hline
 1D$\times$1D$\times$1D & $(1+q_{1}^{2}\xi_{1}^{2})^{-1}(1+q_{2}^{2}\xi_{2}^{2})^{-1}(1+q_{3}^{2}\xi_{3}^{2})^{-1}$ \\
 2D$\times$1D & $(1+q_{1}^{2}\xi_{1}^{2}+q_{2}^{2}\xi_{2}^{2})^{-1.5}(1+q_{3}^{2}\xi_{3}^{2})^{-1}$\\
 3D & $(1+q_{1}^{2}\xi_{1}^{2}+q_{2}^{2}\xi_{2}^{2}+q_{3}^{2}\xi_{3}^{2})^{-2}$
 \end{tabular}
 \end{ruledtabular}
 \end{table}

If, perhaps upon appropriate re-scaling of co-ordinates, the
disorder is isotropic, such as introduced for example by the domain
structure in the random field Ising model (RFIM),
\cite{ZacharZaliznyak} phase slips depend only on $|{\bf R}_{j}|$
and $\langle \phi_{j}^2\rangle/2 = \frac{|{\bf n}_{j}|}{\xi}$. While
the lattice sum can not be easily evaluated, it can be rewritten as
an integral which is repeated periodically in reciprocal lattice and
summed to restore the lattice translational symmetry. For a
$D$-dimensional lattice ($D = 1,2,3$), the result is a
generalized-lattice-Lorentzian function,
\begin{equation}\label{ndim}
   \sum_{\btau} \left(1+\displaystyle\sum_{\alpha=1}^{D}(q_{\alpha}\pm
   Q_{\alpha}+\tau_{\alpha})^2\xi_{\alpha}^{2}\right)^{-\frac{D+1}{2}},
\end{equation}
\noindent where $\xi_{\alpha}$ are the original un-rescaled
correlation lengths and $\btau$ are reciprocal lattice vectors.
Cross-section of the form given by Eq. \ref{ndim} was observed in
neutron scattering experiments in two- and three-dimensional random
field Ising ferro- and antiferro-magnets, in particular in
Rb$_{2}$Co$_{0.7}$Mg$_{0.3}$F$_{4}$ and
Co$_{0.35}$Zn$_{0.65}$F$_{2}$. \cite{Birgeneau1983,Hagen1983}

\begin{figure}[!h]
\includegraphics[width=3.2in,angle=0]{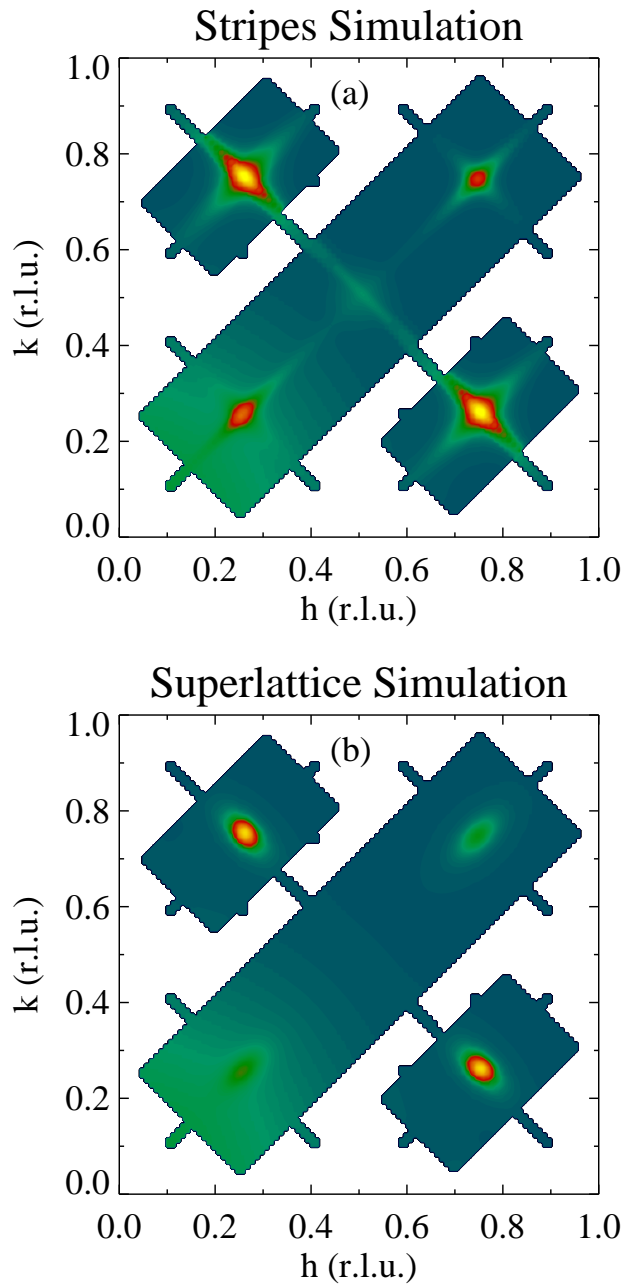}
\caption{\label{sims} Simulated magnetic scattering intensity for
La$_{1.5}$Sr$_{0.5}$CoO$_4$ in the (h,k,0) reciprocal lattice zone
with anisotropic correlation lengths in the a-b plane $\xi_{1,1,0} =
9.2$ and $\xi_{1,-1,0} = 18.4$ LTO (HTT diagonal) lattice units
(l.u.). 2:1 ratio of the correlation lengths corresponds to findings
of Ref. \onlinecite{Du2000}. (a) factorized 1D lattice Lorentzian
cross-section (b) Lorentzian-squared corresponding to anisotropic 3D
disorder.}
\end{figure}

Scattering functions for different combinations of disorder
described by Eqs. (\ref{latticelorentzian}) and (\ref{ndim}) on a 3D
lattice are summarized in Table \ref{coupling}. A fully factorized
(product of LL in all 3 directions) 1D$\times$1D$\times$1D
cross-section can be expected in stripe picture for
La$_{1.5}$Sr$_{0.5}$CoO$_4$. Indeed, discommensurations destroy
magnetic correlation perpendicular to stripes without seriously
affecting order along them. Similarly, the inter-plane correlation
is destroyed by faults in plane stacking. Resulting diffuse
scattering has diamond-like shape reminiscent of a superposition of
quasi-1D "rods" of scattering extended perpendicular to stripes
and/or planes, such as shown in Fig. \ref{sims}(a). Cross-section
corresponding to anisotropic 3D domains given by a
lattice-Lorentzian-squared is shown in Fig. \ref{sims}(b)

Whether the disorder-generating defects are independent
linear/planar disclinations or not (i. e. independent of how the
cross-section is factorized and which of the models listed  in Table
\ref{coupling} is appropriate), one expects an in-plane anisotropy
between the correlation length along and perpendicular to stripe
direction within the stripe model.\cite{Du2000} Hence, within this
model we expect two weak contributions to magnetic scattering in our
sample, at the diagonal positions h = k $\approx 1/4, 3/4$ in the
(h,k,0) zone of the tetragonal unit cell, which are extended along
this diagonal, indicating shorter correlations perpendicular to
stripes. Their intensity is weak because they are just tails of
magnetic peaks at (1/4,1/4,1) and (3/4,3/4,1) resulting from finite
correlation length (peak width) along the $c$ axis. The strong
signal, which is present in our data at (1/4,3/4,0) and (3/4,1/4,0),
arises from twin magnetic domains in the sample, and thus its
intensity pattern is rotated by 90$^\circ$. Overall, all peaks
should exhibit C$_{2}$ symmetry and contributions from twin domains
should be rotated 90$^\circ$ with respect to each other.

In Figure \ref{sims} we show simulated magnetic scattering intensity
for our sample arising from anisotropic short-range magnetic
correlations expected in the stripe model with correlation length
ratio 2:1, chosen to compare with the data of Ref.
\onlinecite{Du2000}. Fig \ref{sims} (a) shows simulated neutron
scattering data for the factorized-lattice-Lorentzian cross-section,
while scattering corresponding to the lattice-Lorentzian-squared
from anisotropic 3D correlations is presented in Fig \ref{sims} (b).
Equal contributions from both twin domains were assumed.

\begin{figure}[!ht]
\includegraphics[width=3.2in,angle=0]{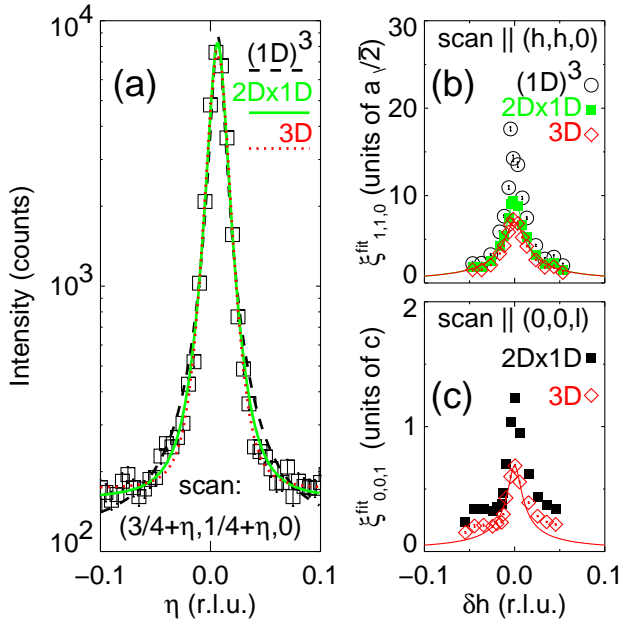}
\caption{\label{xidep}(a) Typical scan through magnetic peak with
fits to cross-sections for completely independent linear
disclinations [(1D)$^{3}$], isotropic disorder in $a-b$ plane with
stacking faults along $c$ (2Dx1D), or disorder coupled in all 3
directions (3D). (b) and (c) ``correlation length'' for scans offset
by $\delta q$ from the magnetic peak position, {\bf Q} $\approx$
(0.744+$\delta h$,0.256-$\delta h$,0) for (b) and {\bf Q} $\approx$
(0.256+$\delta h$,0.256+$\delta h$,1) for (c). The solid/dashed
lines are single-parameter fits to Eq. \ref{Lorfit}, $\delta$q =
$\delta h \sqrt{2}$.}
\end{figure}


\begin{figure*}[!ht]
\includegraphics[width=6.7in,angle=0]{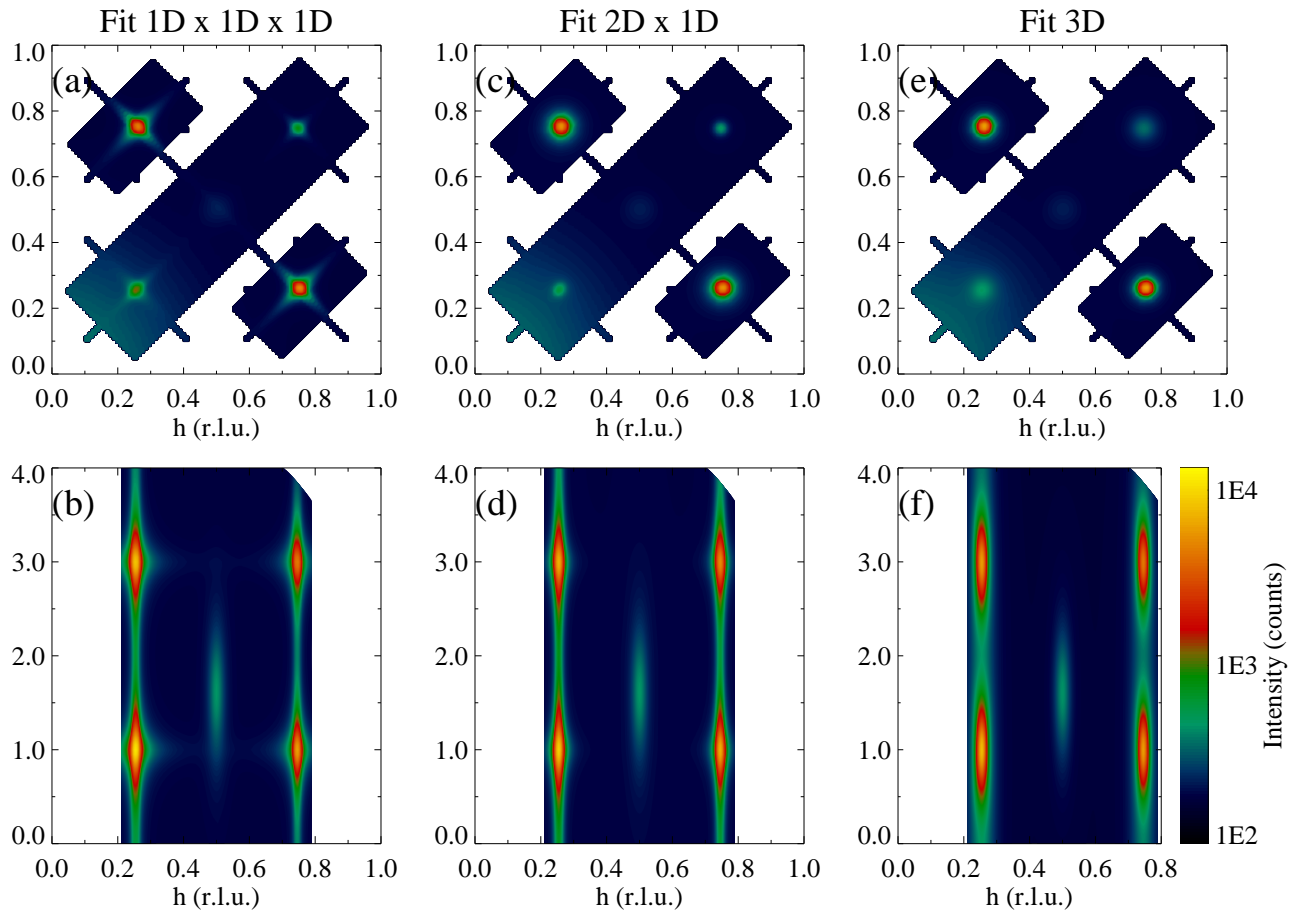}%
\caption{\label{fits} Fit results of neutron scattering intensity in
(h,k,0) (a),(c),(e) and (h,h,l) (b),(d),(f) zones at T=3.5K and 10K.
1D$\times$1D$\times$1D correlation (stripe-like) are in (a) and
(b).(c) and (d) show results for 2D$\times$1D correlations. (e) and
(f)Fitted intensity patterns for coupled isotropic correlations
(superlattice)}
\end{figure*}

From comparing Fig. \ref{data} (a, b) and Fig. \ref{sims} (a, b) it
is already clear that short-range magnetic order in
La$_{1.5}$Sr$_{0.5}$CoO$_4$ is neither anisotropic in the $a-b$
plane, nor it is described by independent one-dimensional magnetic
disclinations associated with stripes running along diagonals of the
HTT unit cell, Fig. \ref{superlattice} (b). It is rather consistent
with anisotropic short-range 3D superlattice model with equal
correlation lengths in the $a-b$ plane. This can be further
quantified by fitting to Eq. (\ref{ndim}) 1D scans made at different
off-sets $\delta q$ from the magnetic peak position along the
diagonal of the HTT unit cell, Fig. \ref{xidep} (a). ``Correlation
lengths'' obtained from such fits of scans along (h,h,0) and (0,0,l)
directions are shown in Fig. \ref{xidep}. For now, we neglect the
instrumental resolution effects which are small compared to much
larger width of magnetic/charge order peaks. In this case, for the
factorized LL scattering cross-section so determined $\xi_\alpha$
should be independent of $\delta q$, while for D = 2,3,
\begin{equation}\label{Lorfit}
    \xi_\alpha^{fit}=\xi_\alpha /\sqrt{1+\sum_{\beta\neq\alpha}(\delta q_\beta \xi_\beta)^{2}},
\end{equation}
\noindent where $\xi_\alpha$ is magnetic correlation length in the
corresponding direction. Fits shown in Fig. \ref{xidep} (b) and (c)
using Equation \ref{Lorfit} yield $\xi_{1,1,0}^{3D}$=7.0 and
$\xi_{1,1,0}^{2D \times 1D}$=9.6 (HTT diagonal) lattice units, and
$\xi_{0,0,1}^{3D}$=0.68  l.u. Note, that these values are obtained
neglecting the resoluntion corrections, and thus represent lower
limits for the corresponding correlation length.

While different fits to 1D scan through magnetic peak shown in Fig.
\ref{xidep} (a) can be hardly distinguished, variation of the fitted
correlation length $\xi_{1,1,0}^{fit}$ with off-set from the peak
position shown in Fig. \ref{xidep} (b) is clearly inconsistent with
the factorized scattering cross-section expected for independent
disclinations associated with stripes in the $a-b$ plane. Moreover,
variation of $\xi_c^{fit}$ obtained from measurements around
(1/4,1/4,1) in the (h,h,l) zone shown in Fig. \ref{xidep} (c) rules
out factorization into a 2D dependence in the $a-b$ plane and a 1D
dependence along $c$-axis, such as arises from independent planar
stacking faults.
Therefore, our results are best described by Eq. (\ref{ndim}) with
D=3 and anisotropic correlation lengths, indicating disorder typical
of an anisotropic 3D random field Ising model.
\cite{ZacharZaliznyak} This is further confirmed quantitatively by
fitting the entire data set to resolution corrected cross sections
from Table \ref{coupling}. Such fits yield $\chi^{2}$ per degree of
freedom values of 6.4 (3D), 10.3 (2D$\times$1D) and 13.6
[(1D)$^{3}$]. For the 3D case (fits are shown in Fig. \ref{scans}
and Fig. \ref{fits} (e, f)) we obtain correlation lengths
$\xi_{1,1,0}=\xi_{1,-1,0}$= 9.4 LTO (HTT diagonal) lattice units
(50.9 \AA) and $\xi_c$= 0.58 l.u. (7.25 \AA). For 2D $\times$ 1D
case, $\xi_{1,1,0}=\xi_{1,-1,0}$= 10.2 l.u. (55.2 \AA), $\xi_c$= 0.9
l.u. (11.3 \AA) and for 1Dx1Dx1D case, $\xi_{1,1,0}=\xi_{1,-1,0}$=
13.3 l.u. (72.1 \AA), $\xi_c$= 1.1 l.u. (13.8 \AA). The latter
compare well with the previous results of Ref.
\onlinecite{Zaliznyak}, although now it is clear from our present
data that a factorized cross-section is not an appropriate model for
magnetic scattering in La$_{1.5}$Sr$_{0.5}$CoO$_4$. Color contour
plots of the calculated intensities corresponding to the above
fitting results are shown in Figure \ref{fits}.

Finally, we also found that magnetic scattering pattern in (hk0)
zone allows us to refine small orthorhombic distortion of the
crystal lattice of about 0.6\% in the $a-b$ plane ($a/b\approx
1.006$).

\section{Summary and conclusions}

In summary, incommensurate magnetic and charge superstructures
observed in hole-doped cuprates, nickelates and cobaltates
La$_{2-x}$Sr$_x$MO$_4$ (M = Cu, Ni, Co, Mn) are often described in
terms of discommensurations in the quasi-regular stacking of charge
lines separating antiferromagnetically ordered stripe domains.
Existence of such faults in stripe stacking has two essential
consequences. First, it renders the super-lattice
incommensurability, which can explain the temperature-dependent
incommensurate magnetism observed in hole-doped nickelates with
$0.25 \lesssim x \lesssim 0.5$. \cite{Kajimoto2001,Ishizaka2003}
Secondly, stacking faults truncate the super-lattice coherence,
resulting in a short-range glassy superstructure, which manifests
itself in experiment by finite-width, diffuse peaks of elastic
scattering in place of Bragg reflections.

Experimental studies of short-range magnetic and/or charge
scattering such as presented in this paper provide an important tool
for investigating spin- and charge-ordered phases and testing
various flavors of stripe models. Our results present strong
evidence that stripe-type superstructure is not at the origin of
incommensurate short-range magnetism in the half-doped cobaltate
La$_{1.5}$Sr$_{0.5}$CoO$_4$. This is not completely unexpected, as
charge order in this material occurs independently of magnetic
order, in a well-insulating state and at much higher temperature.
\cite{Zaliznyak} It is mainly driven by lattice electrostatics and
local spin entropy competing with the crystal field splitting of Co
ion's energy levels. Magnetic incommensurability in this picture can
result from an inhomogeneous exchange modulation induced by CO.
\cite{Zaliznyak2003} The rigidity of quasi-1D charge-stripe
segregation, on the other hand, is rendered by the kinetic energy of
charge hopping, \cite{Tranquada1995,Zaanen1989,Emery1993} which
seems insignificant in our case. Our analysis can be applied to
investigating the relevance of kinetic energy driven segregation of
doped charges into stripes in cuprates and for "diagonal stripe" CO
in other insulating La$_{2-x}$Sr$_x$MO$_4$ oxides. Such studies of
"stripe-ordered" nickelates and cuprates are currently under way.

\begin{acknowledgments}
We thank NIST Center for Neutron Research for hospitality and
J.~Tranquada and M.~H\"{u}cker for discussions. This work was
performed under Contract DE-\-AC02-\-98CH10886, Division of Material
Sciences, US Department of Energy, and utilized facilities supported
in part by the National Science Foundation under Agreement
DMR-0454672.
\end{acknowledgments}


\end{document}